\documentclass[a4paper]{article}

\usepackage[english]{babel}
\usepackage[utf8x]{inputenc}
\usepackage[T1]{fontenc}
\usepackage{float}
\usepackage{caption}
\usepackage{chngcntr}

\usepackage{subcaption}

\usepackage[a4paper,top=3cm,bottom=2cm,left=3cm,right=3cm,marginparwidth=1.75cm]{geometry}

\usepackage{amsmath}
\usepackage{graphicx}
\usepackage[colorinlistoftodos]{todonotes}
\usepackage[colorlinks=true, allcolors=blue]{hyperref}

\title{Survey on Embedding Models for \\ Knowledge Graph and its Applications }
\author{Manita Pote \\ potem@iu.edu}
\date{}

\begin{document}

% \section*{Question 1}

% 1. Write a paper that surveys knowledge representations with graphs
% and the use of deep learning techniques to mine knowledge graphs. In
% your paper, discuss how to apply these techniques to social media data
% graph embeddings. We expect a good survey to include around 30
% references. Like any good survey paper, yours should not simply
% summarize other papers, but should present them in an organized way to
% give a coherent snapshot of work in this area. For instance, many
% survey papers compare and contrast different approaches, pointing out
% strengths and weaknesses. You may want to read some survey papers to
% get a sense for how to do this; survey articles from ACM Computing
% Surveys are a good place to start.

\maketitle

\section*{Abstract}
Knowledge Graph (KG) is a graph based data structure to represent facts of the world where nodes represent real world entities or abstract concept and edges represent relation between the entities.
Graph as representation for knowledge has several drawbacks like data sparsity, computational complexity and manual feature engineering.
Knowledge Graph embedding tackles the drawback by representing entities and relation in low dimensional vector space by capturing the semantic relation between them. 
There are different KG embedding models. 
Here, we discuss translation based and neural network based embedding models which differ based on semantic property, scoring function and architecture they use. 
Further, we discuss application of KG in some domains that use deep learning models and leverage social media data.

\section{Introduction}
Knowledge Base and Knowledge Graphs are the structured representation of facts about the world. 
Knowledge Base (KB) represents the facts in the form of set of triples, \textit{(s, p, o)} where subject, \textit{s} is related to object,\textit{o} by relation, p for example: \textit{(Arnold Schwarzenegger, was a governor of, California)}. 
Here, subjects and objects are name and place entities. 
The entities in KB are concrete entities and abstract concepts of the world. Knowledge graphs(KG) are the graph representation of the set of triples where the nodes represents entities and edges represent relation between entities. Triples in KG are usually represented by \textit{(h,r,t)} where \textit{h} is head entity, \textit{r} is relation edge and \textit{t} is tail entity.
Hence, KG is a heterogeneous directed graph with multiple types of nodes and edge labels.  
Graphs as a data model for knowledge, provide a flexible schema to extend growing nature of data as well as allow to use graph algorithms to query, summarise, reason about the semantics of the terms and gain insight about the domain being described by the graph.

There are different models to represent entities and relations in triples which can be domain specific. 
Resource Description Framework(RDF) model identifies subject, predicate and object as Uniform Resource Identifier(URI). 
URIs provide a way of standardization in representation so that facts from different sources can be combined to extend the existing knowledge base.
RDF Schema is a semantic extension of RDF that has vocabularies for building schema that provides semantic to RDF. 
Property-Centric model and wiki data model are two other models to represent KG. 
In Property-Centric model, nodes and edges are represented as key value pair, also known as property graph. This model is mainly used in graph databases like Neo4j. 
Wiki-data model was developed specifically for Wikipedia which represents node as items and properties and edges as statements. 
The difference between RDF and Wikidata-model is that statements contain much more information like statement record references (links to external sources) and qualifier (contextual information about claims) along with triples.
Example:

\begin{figure}[htp]
    \centering
    \includegraphics[width=5cm]{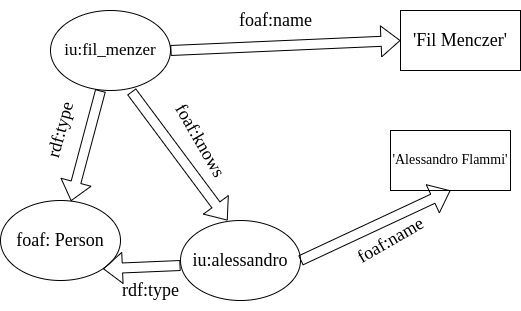}
    \caption{Example of KG representation}
    \label{fig:rdf}
\end{figure}

\begingroup
    \fontsize{5pt}{7pt}\selectfont
    \begin{center}
        \begin{tabular}{ lll } 
         \href{}{https://iu.edu/fil\_menzer} & \href{}{http://www.w3.org/1999/02/22-rdf-syntax-ns\#type} & \href{}{http://xmlns.com/foaf/0.1/Person} \\ 
         \href{}{https://iu.edu/alessandro} & \href{}{http://www.w3.org/1999/02/22-rdf-syntax-ns\#type} & \href{}{http://xmlns.com/foaf/0.1/Person} \\ 
         \href{}{https://iu.edu/fil\_menzer} & \href{}{http://xmlns.com/foaf/0.1/name} & 'Fil Menczer'\\ 
         \href{}{https://iu.edu/alessandro} & \href{}{http://xmlns.com/foaf/0.1/name} & 'Alessandro Flammini'\\
         \href{}{https://iu.edu/fil\_menzer} & \href{}{http://xmlns.com/foaf/0.1/knows} & \href{}{https://iu.edu/alessandro}
        \end{tabular}
    \end{center}
\endgroup
\noindent In the fig.\ref{fig:rdf}, \textit{rdf} and \textit{foaf} prefix represent the vocabularies of RDF. 
Below the figure are the triple representation with URIs of the entities.

\section{Large Scale Knowledge Graphs}
There are several Knowledge Graphs widely used in applications and have great influence. Some of them are given below:

\noindent\textbf{Freebase} ~\cite{freebase} is database of large collection of structured data from many sources like Wikipedia, MusicBrainz, user submitted wiki contributions, etc. 
It has 58,726,427 topics and 3,197,653,841 facts. 

\noindent\textbf{DBpedia} ~\cite{dbpedia} is crowd-sourced community effort to extract structured information from Wikipedia and make this information available on the web. 
It is the largest, best-known and most well maintained knowledge graph. 
It is a multilingual KB that has labels and abstracts for entities in 125 different languages.

\begin{figure}[htp]
    \centering
    \includegraphics[width=8cm]{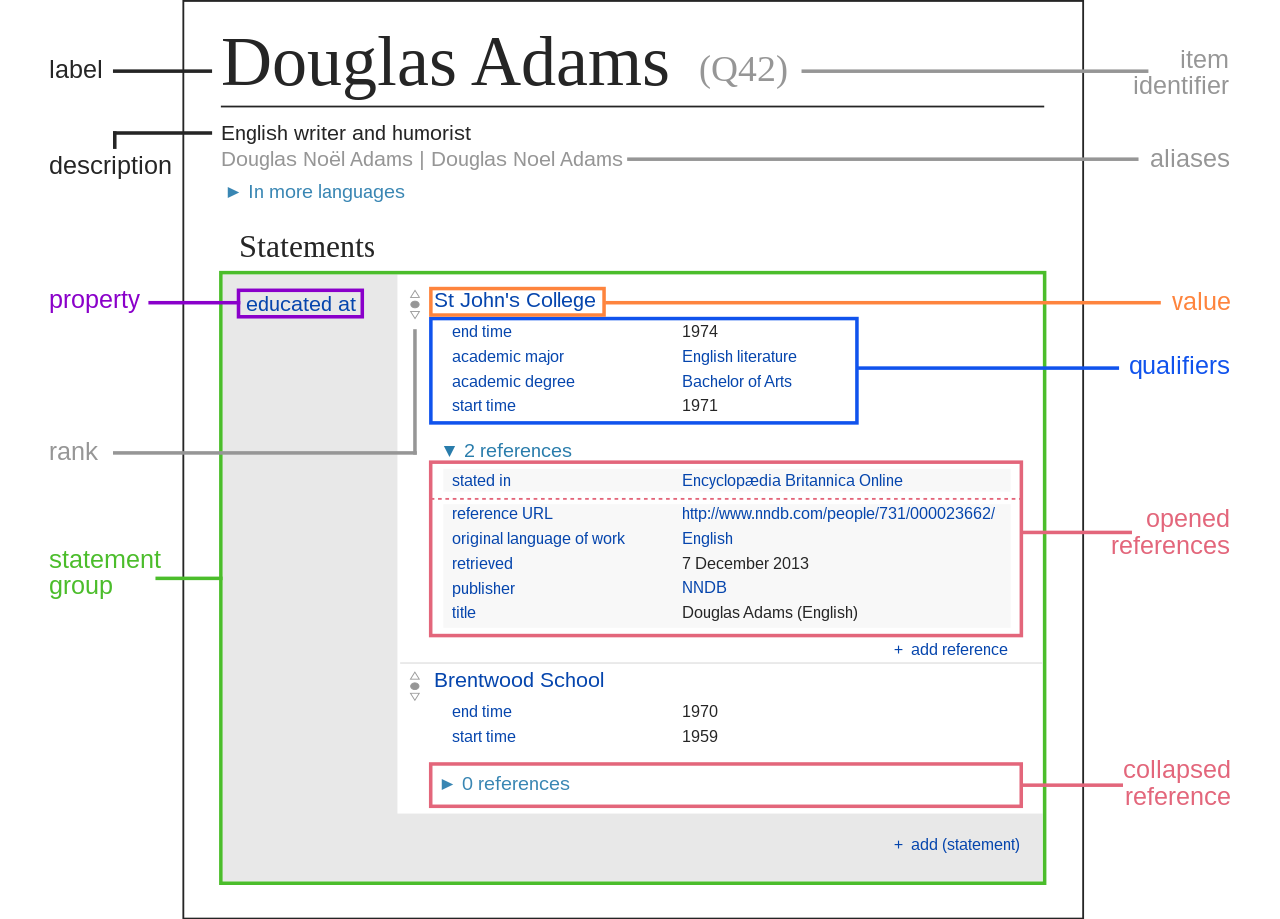}
    \caption{Example of Wiki data representation ~\cite{wikidata}}
    \label{fig:wikidata}
\end{figure}

\noindent \textbf{Wikidata} ~\cite{wikidata} is a free, collaborative, multilingual database collecting structured data to provide support for Wikimedia. 
Wikidata is a document-oriented database, focused on items.
Each item represents a topic and is identified by a unique number. 
Information is added to items by creating statements.
Statements take the form of key-value pairs, with each statement consisting of a property (the key) and value linked to the property.
Example structure of wiki data is given in figure \ref{fig:wikidata}.

\noindent\textbf{YAGO (Yet Another Great Ontology)} ~\cite{yago2008} is a semantic KB derived using automatic extraction methods from three sources: Wikipedia, WordNet and GeoNames. YAGO 4 is the latest version released in 2020.
IBM Watson use YAGO for question answering system.

\section{Deep Learning Models}
Deep Learning is a branch of machine learning which is basically neural network with three or more layers. 
The main difference between traditional machine learning and deep learning is that it can process unstructured data like text and images, automate feature extraction process removing some of the human dependency and gain high accuracy in prediction task. 
The basic processing units in deep learning models are same as neural network: i) input layer ii) hidden layer and iii) output layer. 
There are different deep learning models based on the kind of data they handle and different functions used for computation in each layer. 

Few of the models are described below:
\textbf{Recurrent Neural Network(RNN)}: RNN are a type of feed-forward neural network introduced by Rumelhart \textit{et al.}~\cite{rnn} that can be used to model variable-length sequential information such as sentences or time series.
A basic RNN can be formalized as follows: given input sequence ($x_{1}, ... x_{t}$), for each time step, the model updates the hidden states ($h_1, ... h_t$) and generates the output vector ($o_i, ... o_t$).
The algorithm iterates over the following equations:

\begin{equation}
    \begin{array}{l}
        h_t = tanh(Ux_t + Wh_{t-1} + b) \\
        o_t = Vh_t + c
    \end{array}
\end{equation}

\noindent where U, W and V are the input-to-hidden, hidden-to-hidden and hidden-to-output weight matrices respectively, b and c are bias vectors and $tanh(.)$ is a hyperbolic tangent  non linearity function.\\

\noindent \textbf{Long Short-Term Memory (LSTM)}: In RNN each hidden state is overwritten at each time step. 
LSTM ~\cite{lstm} is a variation of RNN which can maintain memory using memory cell, $c_t$ at time t. 
The output $h_t$ of an LSTM is given by following equations:

\begin{equation}
    \begin{array}{l}
        i_t = \sigma(x_tW_i + h_{t-1}U_i + c_{t-1}V_i) \\
        f_t = \sigma(x_tW_f + h_{t-1}U_f + c_{t-1}V_f) \\
        \overline{c_t} = f_tc_{t-1} + i_t\overline{c_t} \\
        o_t = \sigma(x_tW_o + h_{t-1}U_o + c_tV_o) \\
        h_t = o_t tanh(c_t)
    \end{array}
\end{equation}

\noindent where $\sigma$ is a logistic sigmoid function.
The input gate $i_t$ determines the degree to which the new memory is added to the memory cell. 
The forget gate $f_t$ decides the extent to which the existing memory is forgotten. 
The memory $c_t$ is updated by forgetting part of existing memory and adding new memory $\overline{c_t}$.

\noindent \textbf{Gated Recurrent Unit (GRU)}: 
GRU ~\cite{gru} is similar to LSTM but has gating units that modifies the flow of the content inside the hidden units. 
Following are the equations of GRU layers:

\begin{equation}
    \begin{array}{l}
    z_t = \sigma(x_tU_z + h_{t-1}W_z) \\ 
    r_t = \sigma(x_tU_r + h_{t-1}W_r) \\
    \overline{h_t} = tanh(x_tU_h + (h_{t-1}r_t)W_h) \\
    h_t = (1 - z_t)h_{t-1} + z_t\overline{h_t}
    \end{array}
\end{equation}

\noindent where a reset gate $r_t$ determines how to combine the new input with the previous memory, and an update gate $z_t$ defines how much of the previous memory is cascaded into the current time step and $\overline{h_t}$ denotes the new candidate activation of the hidden state $h_t$.

\begin{figure}[htp]
    \centering
    \includegraphics[width=4cm,height=5cm]{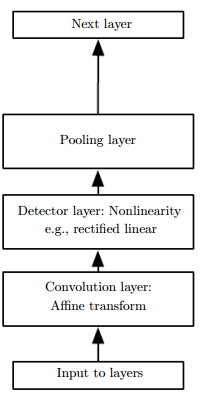}
    \caption{CNN stages ~\cite{goodfellow2016deep}}
    \label{fig:cnn}
\end{figure}

\noindent \textbf{Convolutional Neural Networks:} 
Convolutional neural networks or CNN~\cite{lecun1989backpropagation} ~\cite{goodfellow2016deep} are specialized kind of neural network for processing data that are grid like topology, for example image data. 
Difference between traditional neural network and CNN is that CNN use convolution operation between input data and kernel instead of general matrix multiplication in one of the layer.
The basic stages of the CNN is given in figure \ref{fig:cnn} ~\cite{goodfellow2016deep}. 
In first stage, convolution operation is performed between input and the kernel. 
Kernel is a filter matrix used to extract features from data. 
In second stage output from first stage is passed through non-linearity function and in third stage, a pooling function is used to modify the output of second layer. 
The pooling function replaces the output at certain layer with summary statistic like maximum or mean value within the rectangular neighborhood. 
The motivation behind using a convolution operation is that it allows sparse interaction, parameter sharing and equivariant representation. 
Sparse interaction is accomplished by making kernel smaller than input so that only few parameters are stored. 
Parameter sharing refers to using same parameter for more than one function in a model. 
In a 2D plane data, the kernel with same weights are applied across all the plane which allows parameter sharing and learning fewer weight parameters. Parameter sharing also cause to have a property called equivariance to translation which means it input changes, the output also changes in a same way.

\section{Knowledge Graph Embedding}
Representing information in KG according to its traditional model i.e in the form of network structure has several drawbacks. 
First it is the computational efficiency issue. 
For large volume of data, graph algorithms for mining KG are computationally complex, are not scalable and do not meet the real-time computing needs. 
Second is the data sparsity problem.
Large scale knowledge graph suffers with data sparsity which makes the calculation of semantic or inferential relations of entities extremely inaccurate. 
And third is problem in feature engineering. 
Machine learning task typically use numeric representation or vectors which is different than how KG is expressed therefore machine learning in KG require manual feature engineering~\cite{dai2020survey}.
Knowledge graph embedding tackles the above challenges by representing entities and relations in low-dimensional continuous vector space capturing the semantic relation between them.
The embedding procedure can be described as follows. 
First entities and relations in a given KG are assigned random values of dimension \textit{d}. 
An evaluation function, also called scoring function, is determined to measure the plausibility of triplets.
With each iteration in training, the embedding vectors are updated by maximizing the global plausibility of facts with some optimization algorithm.
The negative examples for the training are generated by replacing the head, \textit{h} and tail \textit{t} in triples by randomly replacing other head and tail entity i.e \textit{(h', r, t)} and \textit{(h, r, t')}. 
A wide range of knowledge graph embedding models have been proposed. 
Here, in this study we summarise the \textit{translation based} and \textit{neural network based} embedding model.

\subsection{Translation-Based Model}
Word2vec~\cite{mikolov2013distributed} is a  word embedding algorithm which represents words in the form of vectors which captures some translation in-variance properties, for example: 

\begin{equation}
\overrightarrow{King} - \overrightarrow{Queen} \approx \overrightarrow{Man} - \overrightarrow{Woman}
\end{equation}

where $\overrightarrow{w}$ is the vector of word \textit{w} obtained from Word2vec model. 
This shows that the word vectors can capture of some of the semantic relationship between words.
Inspired by the Word2vec model, Bordes \textit{et al.}~\cite{bordes2013translating} proposed a TransE model which represents entities (head, \textit{h} and tail, \textit{t}) and relation, \textit{r} in a low dimension vector space, $R^d$ and relation is regarded as connection vector between entities. 
For the true/positive triple (h, r, t), following relation should hold:

\begin{equation}
\vcenter{\hbox{\begin{minipage}{5cm}
    \centering
    \includegraphics[width=7cm,height=5cm]{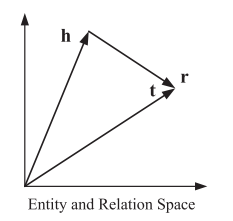}
    \captionof{figure}{Illustration of TransE model~\cite{wang2017knowledge}}
    \end{minipage}
    }
}
\qquad\qquad
    \begin{array}{l}
        h + r \approx t
    \end{array}
\end{equation}

\noindent TransE use $l_2$ as a scoring function and tries to minimize following loss in training set:
\begin{equation}
    \begin{array}{l}
        L = \sum_{(h, r, t) \in S} \sum_{(h', r, t') \in S'} [\gamma + d(h + r, t) - d(h' + r, t')]
    \end{array}
\end{equation}

\noindent where S is sample of positive triplets, S' is sample of negative triplets, $\gamma$ is a margin hyper parameter and $d$ is some dissimilarity measure (eg. Euclidean distance).
TransE model is a simple model but it cannot model complex relation like \textit{symmetry}, for example~\cite{heterogeneouKG} :

If (h, 'Roommate', t) exists in KG then the edge (t, 'Roommate', h) should also exists.
For h, t if (h, r, t) and (t, r, h) holds true then this means
$||h + r - t|| = 0$ and $|| t + r - h|| = 0$

then, $r = 0$ and $h = t$ but $h \text{ and } t$ are two different entities and should be mapped to different location.

Similarly, TransE can not model 1 to N relation, for example~\cite{heterogeneouKG} : If $(h, r, t_1)$ and $(h, r, t_2)$ both exist in KG, e.g., \textit{r} is 'StudentOf' then $t_1$ and $t_2$ will map to same vector although they are different entities.

\begin{equation}
    \begin{array}{l}
        t_1 = h + r = t_2 \\
        t_1 \neq t_2
    \end{array}
\end{equation}

\noindent TransE models embeds relation in same embedding space as the entities.
TransR~\cite{lin2015learning} is model which embeds entities as vectors in the entity space $R^d$ and model each relation as vector in relation space $r \in R^k$ with $M_r \in R^{kxd}$ as the projection matrix.

\begin{equation}
\vcenter{\hbox{\begin{minipage}{5cm}
    \centering
    \includegraphics[width=7cm,height=5cm]{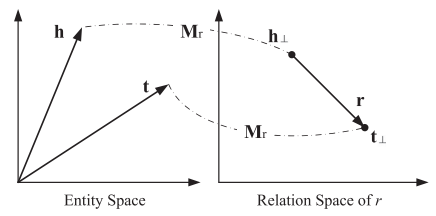}
    \captionof{figure}{Illustration of TransR model~\cite{wang2017knowledge}}
    \end{minipage}
    }
}
\qquad\qquad
    \begin{aligned}
        h_{\bot} = M_rh \\
        t_{\bot} = M_rt
    \end{aligned}
\end{equation}

\noindent where $h_{\bot}$ and $t_{\bot}$ are embedding in relating space. $M_r$ projects entity in space $R^d$ to relation space $R^k$. 
TransR model use following scoring function.

\begin{equation}
    \begin{array}{l}
      f_r(h,t) = -||h_{\bot} + r - t_{\bot}||_{l_2}
    \end{array}
\end{equation}

where $l_2$ is $l_2$-norm. TransR model capture symmetric relation:

\begin{equation}
    \begin{array}{l}
      h_{\bot} = M_rh = M_rt = t_{\bot}, r = 0, h \neq t
    \end{array}
\end{equation}

TransR model also capture 1-to-N relation:

\begin{equation}
    \begin{array}{l}
      t_{\bot} = M_rt_1 = M_rt_2 , t_1 \neq t_2
    \end{array}
\end{equation}

where $t_1$ and $t_2$ are two different entities related to same relation $r$.
However, TransR can not model composition relation. 
Composition relation is defined as follows:

\begin{equation}
    \begin{array}{l}
      (x, r_1, y) \wedge (y, r_2, z) => (x, r_3, z) \forall x, y, z\\
      \text{Example: My mother's husband is my father.~\cite{heterogeneouKG}}
    \end{array}
\end{equation}

\begin{figure}[htp]
    \centering
    \includegraphics[width=7cm,height=5cm]{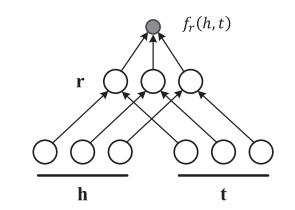}
    \caption{Illustration of DisMult model~\cite{wang2017knowledge}}
    \label{fig:distmult}
\end{figure}

\noindent
DistMult~\cite{DistMult} is a model proposed by Yang \textit{et al.} which use bi-linear scoring function instead of $l_2$ norm. 
The model, figure \ref{fig:distmult}, first represents head(\textit{h}), relation(\textit{r}) and tail(\textit{t}) in low dimension vector $\in R^k$ and then calculates the sum of dot product of h, r and t as score.

\begin{equation}
    \begin{array}{l}
        \text{score function}, f_r(h,t) = \sum_{i}h_{i}.r_{i}.t_{i}
    \end{array}
\end{equation}

\noindent The score function can be viewed as a cosine similarity between \textbf{h.r} and \textbf{t}.
DistMult can model 1-to-N and symmetric relation but can not model anti-symmetric, inverse and composition relation,example~\cite{heterogeneouKG}.

Anti-symmetric Relation:

\begin{equation}
    \begin{array}{l}
        \text{} (h, r, t) => \neg (t, r, h) \forall h, t \\
        \text{Example: Hypernym} \\
        \text{For DistMult:} f_r(h, t) = f_r(t,h) = \text{same score}
    \end{array}
\end{equation}
Inverse Relation:
\begin{equation}
    \begin{array}{l}
        (h, r_2, t) => (t, r_1, h) \\
        \text{Example: Advisor, Advisee relations} \\
        \text{For DistMult:} f_{r_2}(h, t) = f_{r_1}(t,h)\\ \text{This means } r_2 = r_1. \\
    \end{array}
\end{equation}
In this case, advisor and advisee relation is not same. \newline \newline
ComplEx ~\cite{ComplEx}  is a model which embeds entities and relations in complex vector space. This model use following scoring function: \newline
\begin{equation}
\vcenter{\hbox{\begin{minipage}{7cm}
    \centering
    \includegraphics[width=7cm,height=5cm]{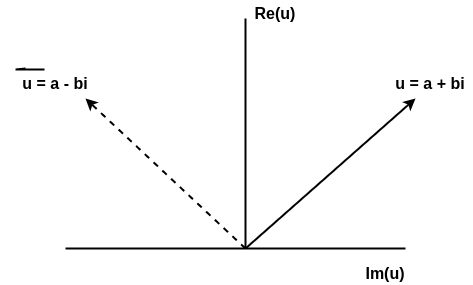}
    \captionof{figure}{Illustration of ComplEx model}
    \label{fig:complx}
    \end{minipage}
    }
}
\qquad\qquad
    \begin{aligned}
         f_r(h, t) = Re(\sum_{i} h_{i}.r_{i}.\overline{t_i})
    \end{aligned}
\end{equation}

where Re is the real part of a complex value and $\overline{t}$ represents the complex conjugate of \textbf{t}. 
ComplEx can model the symmetric, anti symmetric, inverse and 1-to-N relations but not the composition relations.

\subsection{Neural Network Based Models}
Deep Learning is a very popular and extensively used tool in many different fields. 
In recent years, it has been used to embed knowledge graph into continuous feature space using neural network.

\begin{figure}[htp]
    \centering
    \includegraphics[width=6cm,height=5cm]{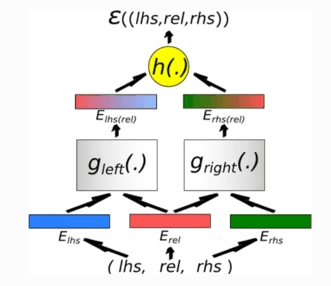}
    \caption{SME model ~\cite{Bordes2014-sme}}
    \label{fig:sme}
\end{figure}

SME ~\cite{Bordes2014-sme} is a neural network based model which uses energy function,$\epsilon$, for the semantic matching to measure the confidence of observed fact (lhs, rel, rhs). 
As show in figure \ref{fig:sme}, SME first maps each input triplet (lhs, rel, rhs) into its embedding, $E_{lhs}, E_{rel} \text{ and } E_{rhs} \in R^d$. 
Then the embeddings, $E_{lhs} \text{ and } E_{rel}$ associated with head, \textit{lhs} and relation, \textit{rel} are used to construct new relation dependent embedding $E_{lhs(rel)}$ for the \textbf{lhs} in the context of relation type represented by $E_{rel}$, similarly for the \textbf{rhs}: $E_{lhs(rel)} = g_{left}(E_{lhs}, E_{rel}) \text{ and } E_{rhs(rel)} = g_{right}(E_{rhs}, E_{rel})$ where $g_{left}$ and $g_{right}$ are parameterized functions tuned during training. 
Finally energy is computed by matching the transformed embeddings in left and right hand side: $\epsilon((lhs,rel,rhs)) = h(E_{lhs(rel)}, E_{rhs(rel)})$ where \textit{h} is either a dot product or other more complex functions whose parameters are learned. 
There are two variations of SME models based whether \textit{g} function is linear or bi-linear.

\noindent - Equations for linear form: \newline
\begin{equation}
    \begin{array}{l}
       E_{lhs(rel)} = g_{left}(E_{lhs}, E_{rel}) = W_{h1}E^T_{lhs} + W_{h2}E^T_{rel} + b^T_h\\
       E_{rhs(rel)} = g_{right}(E_{rhs}, E_{rel}) = W_{t1}E^T_{rhs} + W_{t2}E^T_{rel} + b^T_t\\
    \end{array}
\end{equation}

where $W_{h1}, W_{h2}, W_{t1}, W_{t2} \in R^{pxd}$ (weights), $b_h, b_t \in R^p$ (biases) and $E^T$ denotes the transpose of \textit{E}.

- Equations for bi-linear form:

\begin{equation}
    \begin{array}{l}
        E_{lhs(rel)} = g_{left}(E_{lhs}, E_{rel}) = (W_{h}\overline{x_3}E^T_{rel})E^T_{lhs} + b^T_h\\
       E_{rhs(rel)} = g_{right}(E_{rhs}, E_{rel}) = (W_{t}\overline{x_3}E^T_{rel})E^T_{rhs} + b^T_t\\
    \end{array}
\end{equation}

where $W_h, W_t \in R^{pxdxd}$ (weights) and $b_h, b_t \in R^p$ (biases). $\overline{x_3}$ denotes 3-mode vector.

Energy function:

\begin{equation}
    \begin{array}{l}
        \textbf{$\epsilon$}((h,r,t)) = (g_{left}(E_{lhs}, E_{rel}))^Tg_{right}(E_{rel}, E_{rhs})
    \end{array}
\end{equation}

\noindent SME model can be used for task like link prediction task, predicting whether two entities should be connected by a given relation type, completing missing values of a graph, assessing the quality of a representation etc.

\begin{figure*}[ht]
     \centering
     \begin{subfigure}[b]{0.3\textwidth}
         \centering
         \includegraphics[width=\textwidth]{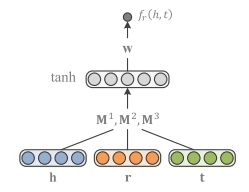}
         \caption{}
         \label{fig:y equals x}
     \end{subfigure}
     \begin{subfigure}[b]{0.3\textwidth}
         \centering
         \includegraphics[width=\textwidth]{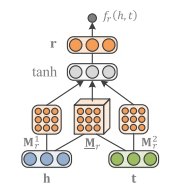}
         \caption{}
         \label{fig:three sin x}
     \end{subfigure}
     \begin{subfigure}[b]{0.3\textwidth}
         \centering
         \includegraphics[width=\textwidth]{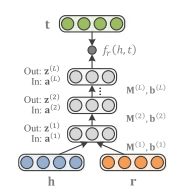}
         \caption{}
         \label{fig:five over x}
     \end{subfigure}
    \caption{Neural Network-based models (a)MLP, (b)NTN, (c)NAM}
    \label{fig:three_graphs}
\end{figure*}

\noindent MLP~\cite{mlps}, figure \ref{fig:three_graphs}, is a simplified light weight multilayer perceptron model which presents relation prediction problem as a  matrix completion task.
KB can be viewed as  a very sparse ${H \times R \times T}$, a 3D matrix G where H, R and T are the number of head, tail and relations. 
Therefore $G(h, r, t) = 1$ if there exists link between h and t else $ G(h, r, t) =  0 $. 
The MLP first maps the h, r, t into its respective embeddings and then calculates the probability of link between h and t for r using the following function:

\begin{equation}
    \begin{array}{l}
        Pr(G(h,r,t) = 1) = w^Ttanh(M^1h + M^2r + M^3t)
    \end{array}
\end{equation}
\noindent where $M^1$, $M^2$, $M^3 \in R^{d \times d}$ are the first layer weights and $w \in R^d$ is the second layer weights shared across different relation.

\begin{figure}[htp]
    \centering
    \includegraphics[width=0.4\textwidth]{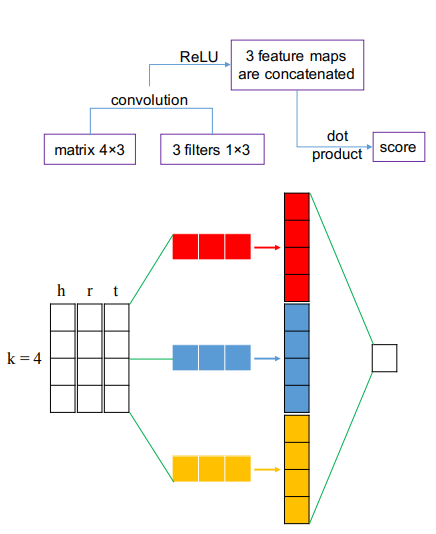}
    \caption{ConvKB model example of embedding size, \textit{k} = 4, number of filters $\tau$ = 3 and activation function, \textit{g}=ReLU  ~\cite{ConvKB}}
    \label{fig:ConvKB}
\end{figure}

\noindent NTN~\cite{ntn}, figure \ref{fig:three_graphs}, is a more generalized neural network architecture that replaces a linear neural network layer with a bi-linear tensor layer that relates two entity vectors across multiple dimensions. 
The model computes how likely it is that two entities are in a certain relationship by the following NTN based function:

\begin{equation}
    \begin{array}{l}
        f_r(h, t) = r^Ttanh(h^TM^{[1:k]}_rt + M^1_rh + M^2_rt + b_r)
    \end{array}
\end{equation}

\noindent where $M^{[1:k]}_{r} \in R^{dxdxk}$ is a tensor and the bi-linear tensor product $h^TM^{[1:k]}_rt$ results in a vector $ y\in R^k$, where each entry is computed one slice $i = 1,... ,k $ of tensor: $y_i = h^TM^{[1:k]}_rt$. The other parameters for relation, $M^1_r, M^2_r$ are standard form of a neural network: $M^1_r, M^2_r \in R^{k\times d}$ and $b_r \in R^k$. $r^T$ is a weight matrix in second layer.

\noindent NAM ~\cite{nam}, figure \ref{fig:three_graphs}, is a deep learning based model for probabilistic reasoning. 
For a given triple, the model first concatenates embedding vector of head entity and relation in the input layer which gives $z^{(0)} = [h;r] \in R^{2d}$. 
The input $z^{(0)}$ is then fed into a deep neural network consisting of \textit{L} rectified linear hidden layers such that

\begin{equation}
    \begin{array}{l}
        a^{l} = M^{(l)}z^{(l-1)} + b^{l}  (l = 1, ....., L) \\
        z^{(l)} = ReLU(a^{(l)}) = max(0, a^{(l)})   (l = 1, .... L)
    \end{array}
\end{equation}

\noindent where $\mathbf{M}^{(l)}$ and $\mathbf{b}^l$ represent the weight matrix and bias for layer \textit{l} respectively.
The score for a given triple is calculated by taking dot product of tail entity and output of last hidden layer, i.e,

\begin{equation}
    \begin{array}{l}
        f_r(h,t) = t^Tz^{(L)}
    \end{array}
\end{equation}

% Neural Tensor Model : The goal is to learn models for common sense reasoning, the ability to realize that some facts hold purely due to other existing relations. (link prediction in an existing network of relationships between entity nodes.) whether ($e_1, e_2$) are the standard form of a neural network: Then association probability for the each triple, $x_n = (e_i, r_k, e_j)$ is calculated using a sigmoid function at the last layer.
% \begin{equation}
%     \begin{array}{l}
%         f(x_n; \Theta) = \sigma(z^{(L)}.v^{(2)}_j)
%     \end{array}
% \end{equation}
% where $\sigma(.)$ is a sigmoid function and \textbf{$\Theta = \{W, V^{(1)}, V^{(2)}, C\}$}

% Similarly, another variation of NAM model is RMNN model which is particularly for multi-relational data. The architecture of RMNN is shown in fig (..). The relation code $c^k$ is connected to all the hidden layers in DNN network.
% \begin{equation}
%     \begin{array}{l}
%         a^{l} = W^{(l)}z^{(l-1)} + B^{l}c^{(k)}  (l = 1, ....., L) \\
%         f(x_n; \Theta) = \sigma(z^{(L)}.v^{(2)}_j + B^{(L+1)}.c^{(k)})
%     \end{array}
% \end{equation}
% where $\sigma(.)$ is a sigmoid function and $\mathbf{\Theta = \{W, B, V^{(1)}, V^{(2)}, C\}}$

\noindent ConvKB ~\cite{ConvKB}, figure \ref{fig:ConvKB}, is convolution neural network(CNN) based model for KB completion task. 
It first represents each triple (h, r, t) as a matrix $\mathbf{A = [v_h, v_r, v_t] \in R^{kx3}}, A_i \in R^{1x3}$ denotes i-th row of \textbf{A}.
A filter, $\mathbf{w} \in R^{nx3}$ is applied  on the convolution layer which generates feature map, $v = [v_i, v_2 ... v_k] \in R^k$ as: \\

\begin{equation}
    \begin{array}{l}
        v_i = g(w.A_i + b)
    \end{array}
\end{equation}
\noindent where $b \in R$ is a bias term and $g$ is some activation function like ReLU.
ConvKB can use different filters to generate feature map. 
If $\Omega \text{ and } \tau$ denotes the set of filters and number of filters i.e $\tau = |\Omega|$. 
The resulting vector after concatenation of feature map would be of dimension $\in R^{{\tau}kx1}$.
A score is computed via a dot product of concatenated feature map with weight vector, $y \in R^{{\tau}kx1}$.
ConvKB score function can be defined as follows:

\begin{equation}
    \begin{array}{l}
        f(h, r, t) = concat(g([v_h, v_r, v_t]*\Omega)).y
    \end{array}
\end{equation}

\noindent where $\Omega$ and \textit{y} are shared parameters, independent of h, r and t; * denote convolution operator and $concat$ denotes concatenation operator.

\noindent Inspired by the popularity of generative adversarial network (GAN), Cai \textit{et al.}~\cite{cai-wang-2018-kbgan} proposed a GAN based model for KB representation, called KBGAN.
Authors argue that KB lacks negative facts and negative examples generated by removing the correct tail entity and randomly sampled from uniform distribution are often completely unrelated and of poor quality. 
Therefore generator in KBGAN generates a set of candidate negative triples with different probabilities.
The general framework of KGBAN is figure \ref{fig:kbgan}:

\begin{figure}
    \centering
    \includegraphics[width=\textwidth]{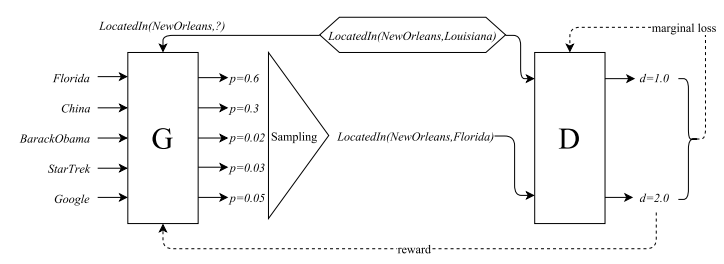}
    \caption{Overview of KBGAN framework ~\cite{cai-wang-2018-kbgan}}
    \label{fig:kbgan}
\end{figure}

\noindent In the figure \ref{fig:kbgan}, the generator (G) calculates a probability distribution over a set of candidate negative triples, then samples one triple from the distribution as output. 
The discriminator (D) receives the generated negative triple as well as the ground truth triples (in the hexagonal box), and calculates the score. G minimizes the score of the generated negative triple by policy gradient and D minimizes the marginal loss between positive and negative triples by gradient descent.

The generator and discriminator can be any probability based KGE model like DistMult, ComplEx, TransE etc.

\begin{figure}[htp]
    \centering
    \includegraphics[width=0.5\textwidth,height=0.4\textwidth]{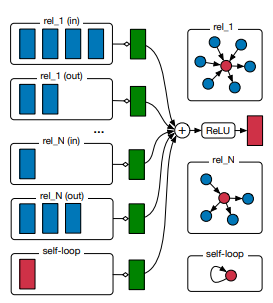}
    \caption{Example of RGCN ~\cite{RGCN}}
    \label{fig:rgcn}
\end{figure}

\noindent Schlichtkrull et al.~\cite{RGCN} proposed a relational graph convolution network model named, R-GCN to handle multi-relational data characteristics of KG, especially for link prediction and node classification task. 
Graph convolution network is based on message-passing framework where computation in each layer is given by following equation:

\begin{equation}
    \begin{array}{l}
        h^{l+1}_i = \sigma(\sum_{m \in M_i}g_m(h^{(l)}_i, h^{(l)}_j))
    \end{array}
\end{equation}

\noindent where $h^{(l)}_i \in R^{d(l)}$ is the hidden state of node $v_i$ in the \textit{l}-th layer with $d^{(l)}$ dimension.
Incoming messages of the from $g_m(.,.)$ are accumulated and passed through an element-wise activation function $\sigma(.)$ $M_i$ denotes the set of incoming messages for node $v_i$. $g_m(.,.)$ are usually message-specific neural network function or simply linear transformation $g_m(h_i, h_j) = Wh_j$ with weight matrix W. 
R-GCN is a special case of this framework where computation in each layer has a relation specific weight matrix and is given by following equation:

\begin{equation}
    \begin{array}{l}
        h^{l+1}_i = \sigma(\sum_{r \in R}\sum_{j \in N^r_i}\frac{1}{c_{i,r}}W^{(l)}_{r}h^{(l)}_j +  W^{(l)}_0h^{(l)}_i)
    \end{array}
\end{equation}

\noindent where $N^r_i$ denotes the set of neighbor indices of node $i$ under relation $r \in R$. $c_{i,r}$ is a problem specific normalization constant that can either be learned or are chosen in advance such as $c_{i,r} = |N^r_{i}|$.
$W_r$ is a relation specific weight matrix. 
$W_0$ is a weight matrix for each node to pass message to subsequent layers (self loop). 
The general framework of R-GCN is given in figure \ref{fig:rgcn}.

\noindent In figure \ref{fig:rgcn}, for each red node/entity, the d-dimensional vector from neighboring nodes (blue rectangles) are gathered and transformed for each relation type individually.
The resulting representation (green) is accumulated in a sum and passed through an activation function. 
If R-GCN is used for entity classification, the softmax function per node is used on the output of last layer.
If it is used in link prediction task, DistMult is used as decoder to score a triple.

\section{Some Application of KG Embedding in NLP:}
Learned entity and relation embedding can be applied to variety of downstream task in NLP like link prediction, triple classification, entity classification, entity resolution etc.

\textbf{Link prediction\:} 
Link prediction refers to task of predicting an entity that has specific relation with another given entity that is predicting \textit{h} given $(r,t)$ or \textit{t} given $(h,r)$.
This is KG completion task where missing knowledge is added to graph. 
Once embedding for entities are learned after training an embedding model, we can rank or score the candidate entities using scoring function. 
For example for (?,r,t) task all the heads in KG can be taken as candidate head and can be scored,$f_r(h',t) = -||h' + r -t||_{1/2}$ if TransE model has been used.

\textbf{Triple Classification:} Triple classification refers to verifying whether an unseen triple fact (h,r,t) is true or not. 
Similar to link prediction task, we can calculate a score for the given triple.
Based on some threshold $\delta_r$ for each relation r, the triple can be classified as true if the score is above the threshold and vice versa.

\textbf{Entity Classification:} The goal of entity classification is to categorize entities into different semantic categories for eg, Fil is a professor and 'The Structure of Scientific Revolution' is a book. 
If 'is a' relation is already encoded in training process of KG, entity classification is a similar task as link prediction i.e \textit{(x, is a , ?)}.

\noindent \textbf{Entity Resolution:} The aim of entity resolution is to verify whether two entities refer to the same object.
In KG, many node can refer to identical objects for example: Arnold Schwarzenegger is an governor of California. 
Schwarzenegger played T-800 in terminator. 
Here, both subject refer to same person and can be defined by a 'equal to' relation. 
Therefore, if 'equal to' relation is encoded in training the model, entity resolution is same task as triple classification.

% \subsection{Named entity recognition}
% \subsection{Entity disambiguation}
% \subsection{Question answering}
% \subsection{Information Extraction}

\section{Application of Knowledge Graph}

 \subsection{Fake News/Rumor Detection}
 With the widespread popularity of Internet and social media, it has become easier to get and disseminate news at low cost without gatekeeper. 
 Along with it, the problem of fake news has also proliferated.
 Fake news are the news articles that are intentionally and verifiably false and could mislead readers~\cite{social-media-fake-news}. 
 Most of the work in fake news detection methods can be categorized as style-based and knowledge-based~\cite{fakenews_content_2018}. 
 Style based methods capture the writing style of news content. 
 Knowledge-Based approach detects fake news is based on the truthfulness of statements claimed on the news content. 
 This is also known as fact checking.  Shiralkar \textit{et al.} ~\cite{finding_stream} takes fact checking as a \'Minimum Cost Maximum Flow Problem\' and propose a fact-checking algorithm called Relational Knowledge Linker that verifies a claim by finding a shortest, semantically related path in existing KG. 
 For semantic similarity, they used cosine similarity of TF-IDF vectors of co-occurrence of nodes. 
 The proposed method does not work well for incomplete graphs and in the case where there is no KG in the first place hence Pan \textit{et al.} ~\cite{fakenews_content_2018} present a KG embedding based fake news detection framework that first extracts triples from summaries of fake and true news articles to build a KG then use variations of TransE model to get embedding of entities and relations and build a SVM classifier using the embedding as features. 
 Ma \textit{et al.}~\cite{fakenews_RNN_2016} first introduce the deep learning model for rumor detection.
 They treat rumors as continuous events thus propose RNN, LSTM and GRU based model for learning the hidden representations that capture the variation of contextual information of relevant Twitter post over time. 
 They use TF-IDF vectors of time binned posts as input for the models.

 \subsection{Drug related application}
 Collecting accurate and up-to-date information regarding drug-use is a difficult task because of the illicit nature of the topic. 
 Tassone \textit{et al.}~\cite{tassone2020utilizing} developed a CNN based classifier to predict whether the tweet is drug related or not. 
 The classifier uses vectors obtained from Word2vec model for each sentences as input.
 The authors further developed a drug relation graph from the predicted drug positive tweets via an associate rule association mining.
 In recent years, social media such as Instagram, has served as a fertile ground for illicit drug trading. 
 To automatically detect illicit drug traffickers on social media, Qian \textit{et al.}~\cite{qian2021distilling} proposed a framework named MetaHG. 
 In the framework, first a heterogeneous graph is constructed with three different types of entities (user, post and drug related keywords) and eight types of relations as well as features of each node.
 The constructed graph is further refined using similarity measures. Then the self supervised R-GCN model is used to get node embedding.
 Finally, the embedding is passed through meta knowledge distillation module to classify the node. 
 
 \subsection{Suicidal Ideation}
 Social media has been an important data source for getting information about suicidal tendencies and other mental health conditions related to depressed users. 
 Gaur \textit{et al.}~\cite{mental-health} proposed a CNN based framework that incorporates domain-specific knowledge as well as data from Reddit to predict the severity of suicide risk for an individual. 
 The authors developed a suicide risk severity lexicon using medical knowledge base and suicide ontology to detect cues relevant to suicidal thoughts and actions. 
 For the social media data they used medical named entity recognition to build a dataset of redditors that have discussed or implied suicidal ideation, behavior or attempt. 
 Cao \textit{et al.}~\cite{personal-knowledge-graph} introduced a new concept of personal knowledge graph for suicidal ideation and proposed a graph attention network based framework for suicidal ideation detection in social media. 
 The personal knowledge graph is constructed from user's personal information like age, gender, location, user experience (stress, psychological disorder, previous suicide attempt), personality, posting behavior and emotion expression.
 
\subsection{Completing Knowledge Graph with Social Media Data}
Knowledge Graph like Wikidata and YAGO are often constructed by incorporating knowledge from semi-structured heterogeneous data sources such as Wikipedia.
However, despite of large amount of knowledge, these graphs are still incomplete. Hazimeh \textit{et al.}~\cite{social-network-profile} proposed a KG refinement method of adding missing knowledge to KG i.e social media profile links for scholars. 
The proposed algorithm works as follows: 1) it takes the name of scholars as query and retrieves its corresponding Wikidata Knowledge graph and its corresponding Google Scholar profile 2) it combines information from these two into one knowledge base 3)it matches this knowledge base with the information crawled from Twitter and Facebook and gets a feature vector based on semantic and syntactic similarity 4) it clusters the feature vectors and filter outs the features below critical threshold and 5) it builds a binary classifier to find out the matched profile.

\section{Conclusion}
In this study, we discuss basics of KG representation, some deep learning models and KG embedding model. 
In particular, translation based and neural network based embedding models that differ based on semantic property they capture, scoring function and architecture. 
KG are usually domain specific. 
Here, we discuss few of the application areas that leverage social media data.

The strength of KB embedding is that it can capture the context specific meaning really well. 
Therefore, KB embedding based models can be useful in domains where the semantic meaning is more important.
Future possible research using KG embedding can be in detecting propaganda in news articles along with news source information and misinformation spreader in social media combining heterogeneous information of user. 
% for example propaganda detection. Propaganda can mainly take three forms based on the source and its accuracy - white, grey and black. White propaganda comes from a source that is identified correctly and the information in the message tends to be accurate. When the source is concealed the propaganda is called black propaganda and grey propaganda is somewhere between white and black propaganda

% \newpage
% \bibliographystyle{plain}
\bibliographystyle{ieeetr}

\bibliography{refs}

\begin{thebibliography}{10}

\bibitem{freebase}
K.~Bollacker, C.~Evans, P.~Paritosh, T.~Sturge, and J.~Taylor, ``Freebase: A
  collaboratively created graph database for structuring human knowledge,'' in
  {\em Proceedings of the 2008 ACM SIGMOD International Conference on
  Management of Data}, SIGMOD '08, (New York, NY, USA), p.~1247–1250,
  Association for Computing Machinery, 2008.

\bibitem{dbpedia}
C.~Bizer, J.~Lehmann, G.~Kobilarov, S.~Auer, C.~Becker, R.~Cyganiak, and
  S.~Hellmann, ``Dbpedia-a crystallization point for the web of data,'' {\em
  Journal of web semantics}, vol.~7, no.~3, pp.~154--165, 2009.

\bibitem{wikidata}
``Wikidata:introduction.''
  \url{https://www.wikidata.org/wiki/Wikidata:Introduction}.
\newblock Accessed: 2022-07-21.

\bibitem{yago2008}
F.~M. Suchanek, G.~Kasneci, and G.~Weikum, ``Yago: A large ontology from
  wikipedia and wordnet,'' {\em Journal of Web Semantics}, vol.~6, no.~3,
  pp.~203--217, 2008.

\bibitem{rnn}
G.~E.~H. David E.~Rumelhart and R.~J. Williams, ``Learning representations by
  back-propagating errors,'' {\em Nature}, vol.~323, pp.~533--536, 10 1986.

\bibitem{lstm}
S.~Hochreiter and J.~Schmidhuber, ``Long short-term memory,'' {\em Neural
  Comput.}, vol.~9, p.~1735–1780, nov 1997.

\bibitem{gru}
K.~Cho, B.~van Merri{\"e}nboer, D.~Bahdanau, and Y.~Bengio, ``On the properties
  of neural machine translation: Encoder{--}decoder approaches,'' in {\em
  Proceedings of {SSST}-8, Eighth Workshop on Syntax, Semantics and Structure
  in Statistical Translation}, (Doha, Qatar), pp.~103--111, Association for
  Computational Linguistics, Oct. 2014.

\bibitem{lecun1989backpropagation}
Y.~LeCun, B.~Boser, J.~S. Denker, D.~Henderson, R.~E. Howard, W.~Hubbard, and
  L.~D. Jackel, ``Backpropagation applied to handwritten zip code
  recognition,'' {\em Neural computation}, vol.~1, no.~4, pp.~541--551, 1989.

\bibitem{goodfellow2016deep}
I.~Goodfellow, Y.~Bengio, and A.~Courville, {\em Deep learning}.
\newblock MIT press, 2016.

\bibitem{dai2020survey}
Y.~Dai, S.~Wang, N.~N. Xiong, and W.~Guo, ``A survey on knowledge graph
  embedding: Approaches, applications and benchmarks,'' {\em Electronics},
  vol.~9, no.~5, p.~750, 2020.

\bibitem{mikolov2013distributed}
T.~Mikolov, I.~Sutskever, K.~Chen, G.~S. Corrado, and J.~Dean, ``Distributed
  representations of words and phrases and their compositionality,'' {\em
  Advances in neural information processing systems}, vol.~26, 2013.

\bibitem{bordes2013translating}
A.~Bordes, N.~Usunier, A.~Garcia-Duran, J.~Weston, and O.~Yakhnenko,
  ``Translating embeddings for modeling multi-relational data,'' {\em Advances
  in neural information processing systems}, vol.~26, 2013.

\bibitem{wang2017knowledge}
Q.~Wang, Z.~Mao, B.~Wang, and L.~Guo, ``Knowledge graph embedding: A survey of
  approaches and applications,'' {\em IEEE Transactions on Knowledge and Data
  Engineering}, vol.~29, no.~12, pp.~2724--2743, 2017.

\bibitem{heterogeneouKG}
J.~Leskovec, ``Heterogeneous graphs and knowledge graph embeddings.'' 2022.

\bibitem{lin2015learning}
Y.~Lin, Z.~Liu, M.~Sun, Y.~Liu, and X.~Zhu, ``Learning entity and relation
  embeddings for knowledge graph completion,'' in {\em Twenty-ninth AAAI
  conference on artificial intelligence}, 2015.

\bibitem{DistMult}
B.~Yang, S.~W.-t. Yih, X.~He, J.~Gao, and L.~Deng, ``Embedding entities and
  relations for learning and inference in knowledge bases,'' in {\em
  Proceedings of the International Conference on Learning Representations
  (ICLR) 2015}, 2015.

\bibitem{ComplEx}
T.~Trouillon, J.~Welbl, S.~Riedel, E.~Gaussier, and G.~Bouchard, ``Complex
  embeddings for simple link prediction,'' in {\em Proceedings of the 33rd
  International Conference on International Conference on Machine Learning -
  Volume 48}, ICML'16, p.~2071–2080, JMLR.org, 2016.

\bibitem{Bordes2014-sme}
A.~Bordes, X.~Glorot, J.~Weston, and Y.~Bengio, ``A semantic matching energy
  function for learning with multi-relational data,'' {\em Mach. Learn.},
  vol.~94, pp.~233--259, Feb. 2014.

\bibitem{mlps}
X.~Dong, E.~Gabrilovich, G.~Heitz, W.~Horn, N.~Lao, K.~Murphy, T.~Strohmann,
  S.~Sun, and W.~Zhang, ``Knowledge vault: A web-scale approach to
  probabilistic knowledge fusion,'' in {\em Proceedings of the 20th ACM SIGKDD
  International Conference on Knowledge Discovery and Data Mining}, KDD '14,
  (New York, NY, USA), p.~601–610, Association for Computing Machinery, 2014.

\bibitem{ConvKB}
T.~D.~N. Dai Quoc~Nguyen, D.~Q. Nguyen, and D.~Phung, ``A novel embedding model
  for knowledge base completion based on convolutional neural network,'' in
  {\em Proceedings of NAACL-HLT}, pp.~327--333, 2018.

\bibitem{ntn}
R.~Socher, D.~Chen, C.~D. Manning, and A.~Y. Ng, ``Reasoning with neural tensor
  networks for knowledge base completion,'' in {\em Proceedings of the 26th
  International Conference on Neural Information Processing Systems - Volume
  1}, NIPS'13, (Red Hook, NY, USA), p.~926–934, Curran Associates Inc., 2013.

\bibitem{nam}
Q.~Liu, H.~Jiang, A.~Evdokimov, Z.-H. Ling, X.~Zhu, S.~Wei, and Y.~Hu,
  ``Probabilistic reasoning via deep learning: Neural association models,''
  {\em arXiv preprint arXiv:1603.07704}, 2016.

\bibitem{cai-wang-2018-kbgan}
L.~Cai and W.~Y. Wang, ``{KBGAN}: Adversarial learning for knowledge graph
  embeddings,'' in {\em Proceedings of the 2018 Conference of the North
  {A}merican Chapter of the Association for Computational Linguistics: Human
  Language Technologies, Volume 1 (Long Papers)}, (New Orleans, Louisiana),
  pp.~1470--1480, Association for Computational Linguistics, June 2018.

\bibitem{RGCN}
M.~Schlichtkrull, T.~N. Kipf, P.~Bloem, R.~v.~d. Berg, I.~Titov, and
  M.~Welling, ``Modeling relational data with graph convolutional networks,''
  in {\em European semantic web conference}, pp.~593--607, Springer, 2018.

\bibitem{social-media-fake-news}
H.~Allcott and M.~Gentzkow, ``Social media and fake news in the 2016
  election,'' {\em Journal of Economic Perspectives}, vol.~31, pp.~211--36, May
  2017.

\bibitem{fakenews_content_2018}
J.~Z. Pan, S.~Pavlova, C.~Li, N.~Li, Y.~Li, and J.~Liu, ``Content {Based}
  {Fake} {News} {Detection} {Using} {Knowledge} {Graphs},'' in {\em The
  {Semantic} {Web} – {ISWC} 2018}, vol.~11136, pp.~669--683, Cham: Springer
  International Publishing, 2018.
\newblock Series Title: Lecture Notes in Computer Science.

\bibitem{finding_stream}
P.~Shiralkar, A.~Flammini, F.~Menczer, and G.~L. Ciampaglia, ``Finding streams
  in knowledge graphs to support fact checking,'' in {\em 2017 IEEE
  International Conference on Data Mining (ICDM)}, pp.~859--864, 2017.

\bibitem{fakenews_RNN_2016}
J.~Ma, W.~Gao, P.~Mitra, S.~Kwon, B.~Jansen, K.~Wong, and M.~Cha, ``Detecting
  rumors from microblogs with recurrent neural networks,'' {\em IJCAI
  International Joint Conference on Artificial Intelligence},
  vol.~2016-January, pp.~3818--3824, 2016.
\newblock 25th International Joint Conference on Artificial Intelligence, IJCAI
  2016 ; Conference date: 09-07-2016 Through 15-07-2016.

\bibitem{tassone2020utilizing}
J.~Tassone, P.~Yan, M.~Simpson, C.~Mendhe, V.~Mago, and S.~Choudhury,
  ``Utilizing deep learning and graph mining to identify drug use on twitter
  data,'' {\em BMC Medical Informatics and Decision Making}, vol.~20, no.~11,
  pp.~1--15, 2020.

\bibitem{qian2021distilling}
Y.~Qian, Y.~Zhang, Y.~Ye, and C.~Zhang, ``Distilling meta knowledge on
  heterogeneous graph for illicit drug trafficker detection on social media,''
  {\em Advances in Neural Information Processing Systems}, vol.~34,
  pp.~26911--26923, 2021.

\bibitem{mental-health}
M.~Gaur, A.~Alambo, J.~P. Sain, U.~Kursuncu, K.~Thirunarayan, R.~Kavuluru,
  A.~Sheth, R.~Welton, and J.~Pathak, ``Knowledge-aware assessment of severity
  of suicide risk for early intervention,'' in {\em The World Wide Web
  Conference}, WWW '19, (New York, NY, USA), p.~514–525, Association for
  Computing Machinery, 2019.

\bibitem{personal-knowledge-graph}
L.~Cao, H.~Zhang, and L.~Feng, ``Building and using personal knowledge graph to
  improve suicidal ideation detection on social media,'' {\em IEEE Transactions
  on Multimedia}, vol.~24, pp.~87--102, 2022.

\bibitem{social-network-profile}
H.~Hazimeh, E.~Mugellini, S.~Ruffieux, O.~A. Khaled, and P.~Cudr\'{e}-Mauroux,
  ``Automatic embedding of social network profile links into knowledge
  graphs,'' in {\em Proceedings of the Ninth International Symposium on
  Information and Communication Technology}, SoICT 2018, (New York, NY, USA),
  p.~16–23, Association for Computing Machinery, 2018.

\end{thebibliography}
\end{document}